\documentclass[conference]{IEEEtran}
\usepackage[letterpaper, top=0.7in, bottom=1in, left=0.625in, right=0.625in]{geometry}
\usepackage{array}
\usepackage{pgfplotstable}
\usepackage{booktabs}
\usepackage{bbm}      
\usepackage{subfig}
\usepackage{graphicx,dblfloatfix}
\usepackage{caption}
\usepackage{color}
\usepackage{amssymb}   
\usepackage{amsmath}
\usepackage{amsthm}
\usepackage{algorithm}
\usepackage{algpseudocode}
\usepackage{enumerate}
\usepackage{tabularx}  
\usepackage{multirow}
\usepackage[draft]{hyperref}      
\usepackage{url} 
\usepackage[T1]{fontenc}
\usepackage{enumerate}
\usepackage{float}
\usepackage{cite}
\usepackage{verbatim}
\usepackage{listings}
\usepackage{setspace}

\begin{document}

\title{A Deep Reinforcement Learning Approach for Adaptive Traffic Routing in Next-gen Networks}% author names and affiliations
\author{
\IEEEauthorblockN{Akshita Abrol\IEEEauthorrefmark{1}, Purnima Murali Mohan\IEEEauthorrefmark{1}, Tram Truong-Huu\IEEEauthorrefmark{1}\IEEEauthorrefmark{2}}
\IEEEauthorblockA{
\IEEEauthorrefmark{1}Singapore Institute of Technology, Singapore \\  
\IEEEauthorrefmark{2}Agency for Science, Technology and Research (A*STAR), Singapore
}
Email: \{akshita.abrol, purnima.mohan, truonghuu.tram\}@singaporetech.edu.sg
}

\maketitle  

\begin{abstract}
Next-gen networks require significant evolution of management to enable automation and adaptively adjust
network configuration based on traffic dynamics. The advent of software-defined networking (SDN) and
programmable switches enables flexibility and programmability. However, traditional techniques that decide traffic policies are usually based on hand-crafted programming optimization and
heuristic algorithms. These techniques make non-realistic assumptions, e.g., considering static network load and topology, to obtain tractable solutions, which are inadequate for next-gen networks. In this paper, we design and develop a deep reinforcement learning (DRL) approach for adaptive traffic routing. We design a deep graph convolutional neural network (DGCNN) integrated into the DRL framework to learn the traffic behavior from not only the network topology but also link and node attributes. We adopt the Deep Q-Learning technique to train the DGCNN model in the DRL framework without the need for a labeled training dataset, enabling the framework to quickly adapt to traffic dynamics. The model leverages $q$-value estimates to select the routing path for every traffic flow request, balancing exploration and exploitation. We perform extensive experiments with various traffic patterns and compare the performance of the proposed approach with the Open Shortest Path First (OSPF) protocol. The experimental results show the effectiveness and adaptiveness of the proposed framework by increasing the network throughput by up to $7.8\%$ and reducing the traffic delay by up to $16.1\%$ compared to OSPF.
\end{abstract}


\section{Introduction}
In the last decade, next-generation (in short: next-gen) networking technologies like software-defined networks (SDN), programmable network devices (e.g., Intel Tofino~\cite{intel}), and network function virtualization (NFV)~\cite{nvidia} have transformed the network into an end-to-end programmable platform. Users can customize networks to their needs and budget. 5G and Beyond networks, with their focus on reliability, low latency, Quality of Services, network slicing, and the Internet of Things, will boost the adoption of these technologies. %On the business side, enterprises face pressure to create revenue-generating dynamic network applications while keeping costs low, necessitating elastic infrastructures for Industry 4.0, intelligent mobility, time-critical IoT, HD video, AR/VR, etc.

%Networked applications utilize various access technologies, including cellular (5G/6G), WiFi (WiFi 6/7), IoT protocols, and satellite (e.g., Starlink), alongside wired networks like time-sensitive networks.

Traffic routing is a crucial aspect of maintaining an efficient network. Traditional routing techniques make it difficult to deal with huge traffic growth and dynamic user demands on today’s Internet~\cite{jiang2021machine}. For instance, OSPF consistently opts for the shortest path regardless of prevailing network conditions. Such inefficient routing decisions will increase transmission delay, and even cause network congestion or packet loss. To address this issue, adaptive traffic routing using artificial intelligence (AI) has been explored recently. AI algorithms are typically categorized into four main groups: supervised learning, unsupervised learning, semi-supervised learning, and reinforcement learning (RL). Supervised learning has been used for various tasks related to traffic routing~\cite{tram2019ml}. However, it is challenging (not to mention impossible) to obtain a high-quality labeled dataset for training supervised machine learning models due to the high velocity of network traffic and difficulty in feature extraction. This also leads to the problem that supervised learning needs a longer time to adapt to traffic dynamics as it needs to capture sufficient labeled data and retrain models in an offline manner. 

In~\cite{sutton2018reinforcement}, the authors suggested an optimal solution for routing by solving its Markov Decision Process (MDP) that iterates over all MDP’s states until convergence. The MDP for the traffic demand allocation problem encompasses all conceivable network topology states and the associated transition probabilities that describe how the system moves from one state to another. However, as the problem size grows, so does the MDP’s state space, making it infeasible for large and complex network topologies. Furthermore, the requirement to specify a discrete state space and the transition probability matrix can sometimes make the solution impractical or less viable. 

Deep Reinforcement Learning (DRL) has demonstrated substantial advancements in addressing challenges related to sequential decision-making and automated control~\cite{valadarsky2017learning}. Consequently, the field of networking is actively exploring DRL as a pivotal technology for enhancing network optimization tasks such as routing, ultimately aiming to facilitate the emergence of adaptively routing networks~\cite{wang2017machine}. It is worth mentioning that computer networks are essentially represented as graphs. Nevertheless, contemporary approaches rely on conventional neural network architectures (such as fully-connected and convolutional neural networks), which are not ideally suited for effectively learning graph-structured data. Graph Neural Networks (GNNs)~\cite{zhou2018graph} represent a unique category of neural networks tailored for scenarios where data is organized in graph structures. GNNs exhibit strong generalization abilities across various graph types and prove valuable in uncovering connections between distinct nodes and edges. These attributes position GNNs as promising candidates for integration into routing algorithms. The inherent generalization power of GNNs empowers routing algorithms to adapt seamlessly to diverse topologies, traffic patterns, and other related factors.

In this paper, we present an adaptive traffic routing approach using deep reinforcement learning. The framework incorporates predictions from deep graph convolutional neural networks (DGCNN) to find the optimal path for traffic flows. The neural networks are trained using past experiences learned in the DRL framework by crafting an appropriate reward function without the need for a labeled dataset. This allows the DGCNN model to adapt to the changes in traffic behavior in real-time and avoid offline model training. We design a node-level feature vector that, combined with the network topology, allows DGCNN to capture meaningful information about the relations between node statistics and traffic behavior. We implement the proposed routing technique using the OpenAI Gym framework and evaluate its performance over two different topologies with various traffic patterns. We compare its performance with that of the Open Shortest Path First (OSPF) and existing RL-based baseline techniques. The proposed approach can be integrated into an SDN controller that is responsible for collecting traffic requests and network data and making routing decisions.

The rest of this paper is organized as follows. Section~\ref{sec:relatedwork} reviews the existing works. The system model and proposed framework are demonstrated in Section~\ref{sec:technique}. Section~\ref{sec:experiments} presents extensive experiments and performance evaluation. Finally, Section~\ref{sec:conclusion} concludes the paper.

\section{Related Work}
\label{sec:relatedwork}

To the best of our knowledge, this is the first work that proposes DGCNN-based adaptive traffic routing using RL setup. However, the use of AI for network optimization has been of great interest among researchers lately. There exist several works such as~\cite{valadarsky2017learning} that use both supervised machine learning and DRL for traffic routing to minimize link over-utilization and demonstrate the effectiveness of the approach. Similarly, the authors of~\cite{xu2018experience} proposed a DRL framework with deep neural networks for traffic engineering and demonstrated its performance in \texttt{ns-3}, a discrete-event network simulator for Internet systems. In~\cite{abbasloo2020classic}, the authors used DRL for congestion control in a specific range of values and derived the best solution from that range, which might be a local optimum rather than the desired global optimum. In~\cite{lan2019deep}, the authors presented a congestion control scheme for named data networks based on DRL that calculates the proper size of the congestion window at the consumer side using the state of variables that are manually selected with no generalization to different topologies. 
In~\cite{emara2020eagle}, the authors leveraged  Bottleneck Bandwidth and Round-trip propagation time (BBR) to train a Long Short-Term Memory model in a DRL agent while generalizing to different network environments and comparing the steady-state behaviors. The learned actions are from simulated environments. In~\cite{roshdi2021deep}, the authors investigated the performance of a decentralized DRL agent in a vehicular network scenario. In~\cite{chen2021ran}, the authors incorporated Radio Access Network (RAN) information to dynamically learn and update the network state. 

Recently, in~\cite{emara2022pareto}, the authors used RL for fairness using multi-agent learning adapting to different network scenarios. In~\cite{almasan2022deep}, the authors developed a DRL framework using message-passing neural networks to address the routing problem in optical circuit-switching networks. It uses only link-level features to maximize traffic volume without any consideration of delay or congestion minimization. The work presented in~\cite{bhavanasi2022routing} approaches traffic routing as an RL problem and utilizes graph convolutional networks; nonetheless, they neglect the incorporation of network features in their model formulation. In~\cite{yang2022joint}, the authors optimized traffic routing and scheduling for time-sensitive networking (TSN) by integrating graph convolutional networks into an RL framework. However, this approach overlooks network throughput optimization. In~\cite{huang2022deep}, the authors employed a deep graph RL model in software-defined wireless sensor networks (SDWSNs) to capture traffic behavior. However, the work solely relies on node-level features and does not consider link throughput, which results in slow convergence. These studies collectively highlight various gaps in existing approaches, pertaining to the consideration of comprehensive network features, optimization objectives, and framework design. Therefore, there remains a need for an adaptive framework that addresses these limitations and provides a holistic approach to traffic routing optimization in network scenarios.

\section{System Design and Algorithm}
\label{sec:technique}

%In this section, we first present the optimization formulation of the adaptive routing problem using MDP and then present our DRL framework, which addresses the limitations of MDP while achieving desirable performance.

\subsection{Formulation of Adaptive Traffic Routing using MDP}

An MDP is defined as a four-tuple $(S, A, R, P)$ where $S$ is the state space, $A$ is the action space, $R$ is the reward function and $P$ is a matrix representing the transition probability from one state to another. Given the current state of the environment (denoted as $s \in S$), the objective of the MDP is to determine an action (denoted as $a \in A$) such that the expected reward is maximized in the long run. Applying action $a$ to the environment (i.e., the network in our problem) makes the environment transition to a new state, denoted as $s'$. 

\subsubsection{State Space}  

To capture the best visibility of the network state, we propose to capture the link state and node state. Combined with the network topology, these form a graph to train the DRL agent in our framework. We provide a detailed description of the link state and node state below.  

\paragraph{Link State} 

It represents up-to-date network topology to avoid faulty or disconnected links. The link state is represented in the form of a Network Adjacency Matrix (NAM) by weights $w_{i,j}$, where $i$ is the source node and $j$ is the target node. The weights have a value of one only if a link is present between the source and the target node, otherwise zero. We add self-loops to NAM to account for the node state. Furthermore, symmetric normalization of NAM is done to ensure that the graph convolution operation is symmetric, which helps capture both local and global features in a more balanced way and improves representational learning. The network state is defined as follows:
  \begin{equation}
  \label{eq:1}
  S^\text{link} = D^{-1/2}W^{'}D^{-1/2}
  \end{equation}
  where $W = \left([w_{i,j}]\right)$, $W^{'} = W + I$ with $I$ be the identity matrix and $D$ is the diagonal node degree matrix of $W^{'}$.
  
  \paragraph{Node State} 
  
  The node state is represented by a feature vector consisting of seven features. For node $i$ in the network, its feature vector is defined as $s_i = [D_{i},I_{i},tx_{i},rx_{i},src_{i},dest_{i},bw_{i}]$ where $D_i$, $I_i$, $tx_i$, $rx_i$, $src_i$, $dest_i$ and $bw_i$ represent the node degree, node importance, aggregated egress rate, aggregated ingress rate, source nodes having traffic destined for node $i$, destination nodes that node $i$ sends the traffic to, and requested data rate, respectively. Both $src_{i}$ and $dest_{i}$ are binary vectors in which a value of one indicates the presence of traffic between node $i$ and the respective node.
  For a network of $N$ nodes, the node state is represented as $S^\text{node} = (s_1, s_2, s_3, ..., s_N)$.
  
It is worth mentioning that node degree indicates multiple path options for higher-degree nodes and vice-versa. Node importance is a measure of centrality inherited from graph theory. It indicates the number of shortest paths that traverse a particular node. In particular, for each pair of nodes in the topology, we compute $k$ (e.g., three) shortest paths and maintain a per-node counter that indicates how many paths pass through it. Ingress and egress rates indicate the amount of traffic load on the nodes, enabling the model to learn paths with less load. The last three features provide the context of the request by the environment. All the values in the feature vector are normalized. Combining $S^\text{link}$ and $S^\text{node}$ forms the Network State Matrix (NSM) for the problem: $s\in S | s = (S^\text{link}, S^\text{node})$.
  
\begin{comment}
\begin{table} 
  \caption{Node Features for NSM.}
  \label{tab:features}
  \begin{center}
  \begin{tabular}{cc}
   \toprule
   Notation & Description \\
   \midrule
    $D_{N}$ & Node Degree \\
    $I_{N}$ & Node Importance \\
    $tx_{N}$ & Bits Transmitted by each node \\
    $rx_{N}$ & Bits Received by each node \\
    $src_{N}$ & One-hot-encoder for Source \\
    $dest_{N}$ & One-hot-encoder for Destination \\
    $bw_{N}$ & Request Datarate \\
    $Z_{N}$ & Zero Padding \\
   \bottomrule
   \end{tabular}
   \end{center}
\end{table}
\end{comment}    
  
\subsubsection{Action Space} 

In the proposed framework, actions represent the probability of choosing a path for a traffic request.  Let $K$ denote the number of possible shortest paths between a pair of source and destination nodes. The action for traffic flow $f$ is defined as
     $a_{f} = (P_{f,1},P_{f,2},\dots, P_{f,K})$
such that $\sum_{k=1}^K P_{f,k} = 1$. The higher the value of $K$, the higher the complexity of the problem. All the shortest paths are computed a priori based on the number of hops or other metrics.
  
\subsubsection{Reward Function} 
  
  The reward symbolizes the overall utility of sending data through a particular path. For our problem, the reward function is modeled to maximize the throughput (transmission rate) and minimize the delay for each traffic flow. It is represented as:
  \begin{equation}
     R = T^{\text{act}}/T^{\text{req}} + 1/D 
     \label{eq:reward}
  \end{equation}
  where $T^\text{act}$ and $D$ represent the actual throughput (available on the path) and delay for each pair of source and destination nodes, respectively. We normalize the actual throughput by dividing it by the requested transmission rate of that flow, $T^{\text{req}}$. The intuition behind the aforementioned formula lies in the fact that the reward would be lower if the path with congestion is chosen given an observed state. Eventually, the agent will be able to learn from those undesired actions with lower rewards as well as better paths during the exploration phase.

\subsubsection{Transition Probability} 
Given the current network state ($s$) and a traffic flow request $f$, applying action $a_f$ and allowing traffic to stream from the source node to the destination node make the network move from state $s$ to state $s'$ with a probability $P(s'|(s, a_f))$. However, it is challenging (not to mention impossible) to derive a closed-form formula to compute this probability. Thus, we leverage reinforcement learning with a model-free approach to solve the adaptive traffic routing problem while achieving an acceptable performance.

   \begin{figure*}[t]
   \centering
   \includegraphics[width=0.85\textwidth]{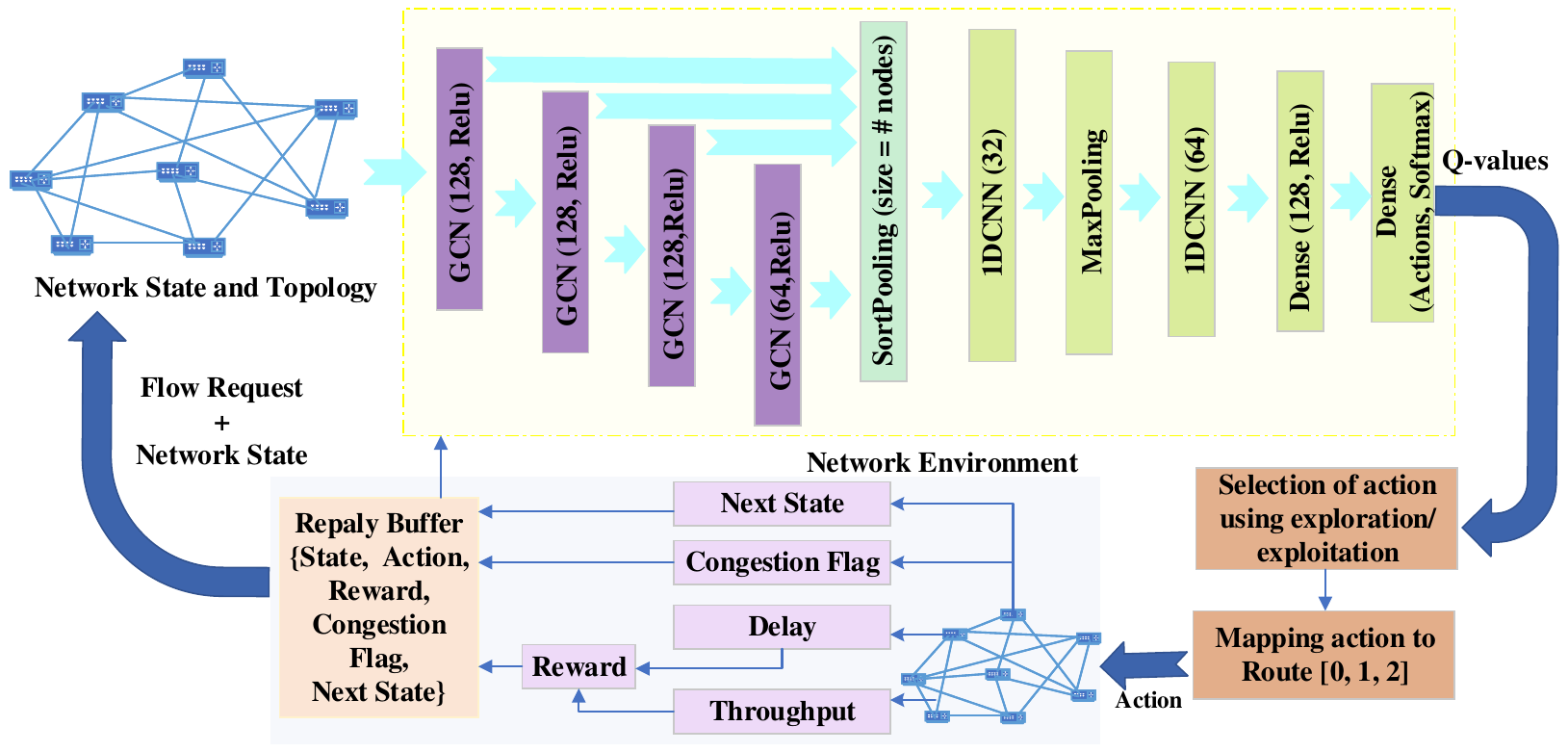} 
   \caption{Overview of Deep Reinforcement Learning  Framework for Adaptive Traffic Routing.}
   \label{fig:system_architecture}
   \vspace{-2ex}
\end{figure*}

%DGCNN has demonstrated its capability in cybersecurity~\cite{pham2021mappgraph}.

\subsection{DRL-based Technique for Adaptive Traffic Routing}
\subsubsection{Deep Graph Convolutional Network Architecture}
DRL-based technique adopts a model-free approach by employing neural networks that can learn the environment (network) state and make decisions on the action to enforce to the network such that the expected reward is maximized in the long run. Depending on the design of the network state (whether it is one-dimensional, multi-dimensional vectors, or graph-based features), the neural networks used in the DRL framework are designed accordingly. With the design of NSM presented in the previous section, we advocate the use of deep graph convolutional neural networks (DGCNN) for our DRL framework. The DGCNN architecture consists of three stages: graph convolution, pooling, and traditional convolution layers followed by dense layers for path prediction as shown in Fig.~\ref{fig:system_architecture}. The input of DGCNN is the network state defined above. In the graph convolution stage, the model computes the node latent representations by propagating information between neighboring nodes through graph convolution layers. This process is performed for multiple layers of sizes $[128,128,128,64]$ with ReLU activation to obtain node representations that capture information from the entire graph. The size of the hidden layers is typically larger than the number of features in the node state. This allows each node to store its features and the aggregated features coming from all the neighboring nodes. 

In the pooling stage, the goal is to combine node representations into a vector with a predefined order and size. Various pooling methods exist, including global pooling and hierarchical pooling. In this work, we use the SortPooling algorithm introduced in~\cite{zhang2018endtoend} to arrange nodes based on the sum of their features at the final layer of the graph convolution stage. When two nodes happen to have identical values at the final layer, the algorithm resolves this by considering the sum of their features at the previous layer until the tie is resolved. We set the size of the pooling layer to the number of nodes of the graph to avoid any loss of information. Finally, the graph represented by a latent vector, is given as input for further learning of local patterns to one-dimensional convolution layers $[32,64]$, MaxPooling layers, fully-connected layer $[128]$, and a \texttt{softmax} layer for prediction of $q$-values, which in turn are converted to the paths to enforce to the network.

\subsubsection{DRL Training} 

The core of the proposed DRL framework is an agent, which runs a DRL algorithm to find the best action at each decision instant, i.e., when there is a traffic flow request. The overview of the DRL framework is shown in Fig.~\ref{fig:system_architecture}. Our framework includes a network environment that defines the optimization problem, encompassing the network topology and node features. Practically, an SDN controller can be deployed to capture the network topology and traffic features such as throughput and delay, forming network states to forward to the DRL agent for path prediction, calculating the reward, and training the DGCNN model. 

In Algorithm~\ref{alg:dqn}, we present the pseudocode of the training process of DRL-GCNN adopted from the Deep Q-Learning algorithm~\cite{mnih2013playing}, which maintains two DGCNN models $Q$ and $Q^{'}$ having the same architectures but different weights initialized by $\theta$ and $\theta^{'}$, respectively. The training process is an infinite loop so that the models are updated whenever there are changes in traffic patterns. At the time instant $t$, the DRL agent receives a traffic flow request $f_t$ defined by the tuple (\texttt{src}, \texttt{dest}, $T^\text{req}$). Based on the network state at time $t$ (denoted as $s_t$), the agent uses the main model ($Q$) to calculate the path for traffic flow $f_t$, denoted as action $a_{f_t}$. Enforcing $a_{f_t}$ to the network for flow $f_t$ makes the network transit to state $s_{t+1}$ and incurs a reward computed using Eq.~\eqref{eq:reward}, denoted as $R_t$. 

It is worth mentioning that action $a_{f_t}$ is calculated using the exploration-exploitation trade-off. Exploration plays a vital role in the training of a DRL agent. This is because a novice agent requires exposure to a significant number of transition samples to acquire expertise and, ideally, develop an optimal policy. For traffic routing problems, exploration becomes particularly formidable due to the presence of a large number of states as well as actions that can be selected during each decision-making step. Hence, we propose to use $\epsilon\text{-greedy exploration}$ in which the policy is defined as follows:
\begin{equation}
 \pi(a_{f_t}|s_t) =
\begin{cases}
     \arg \max_{a_{f_t}}\, Q_t(s_{t}, a_{f_t}, \theta_t) & \text{If } \epsilon < \zeta  \\
     \text{Explore} & \text{Otherwise}.
\end{cases}
\label{eq:cal_policy}
\end{equation}
For greedy exploration, the initial value of $\epsilon = 1$ and $\zeta$ is a randomly generated number uniformly distributed between $0$ and $1$. $\epsilon$ decays exponentially every
episode with rate of $0.99995$ and the minimum value of $\epsilon$ is set to $0.01$. We further use a threshold value $\tau$ defined as the mean of the last $100$ rewards when the main model ($Q$) is in the exploitation state. As, the network traffic behavior changes, the reward value fluctuates outside the range $(\tau-1,\tau+1)$, $\epsilon$ is again set to 1 to enable the model to explore the environment and adaptively update $Q$ and $Q^{'}$. We use a congestion flag (denoted as $c_{t}$) indicating if the allocated path ($a_{f_t}$) exceeds the maximum link capacity. The congestion flag is set as follows:
\begin{equation}
 c_{t} =
\begin{cases}
     1 & \text{if congestion}  \\
     0 & \text{otherwise.}
\end{cases}
\end{equation}
To train the main DGCNN model ($Q$) using stochastic gradient descent, we maintain an experience replay buffer $B$. However, the application of Deep Q-Learning to the routing problem using a simple buffer does not yield satisfactory results. Therefore, we propose to use a prioritized buffer replay, akin to its use in game-playing tasks, to enhance the training process by storing the significance of transition samples. The experience buffer stores $10000$ samples and is implemented as a First-In-First-Out queue. The priority of flow $f_t$, denoted as $p_{f_t}$ is calculated using the Temporal Difference (TD) error $\delta$ during the training process as follows:
 \begin{equation}
     \delta_t = y_{t} -  Q_t(s_t,a_{f_t}, \theta_t); \quad p_{f_t} = \left|\delta_t\right| + \beta
     \label{eq:priority}
  \end{equation}
where $y_{t}$ is the $q$-value of the target model ($Q^{'}$) at time instant $t$, $Q_t(s_{t}, a_{f_{t}}, \theta_t)$ is the $q$-value of the main model ($Q$) at time instant $t$, and $\beta = 0.01$ is used to prevent priority being zero. The $q$-value of the target model ($Q^{'}$) is calculated using the Bellman equation defined as follows:
\begin{equation}
 y_{t} = R_t + (1-c_{t})\gamma\displaystyle\max_{a_{f_{t+1}}}Q^{'}_{t+1}(s_{t+1},a_{f_{t+1}},\theta_{t+1}^{'})
    \label{eq:target}
\end{equation}
where $Q_{t+1}^{'}(s_{t+1},a_{f_{t+1}},\theta_{t+1}^{'})$ is the $q$-value of the target model ($Q^{'}$) at time instant $t+1$, $R_t$ is the reward defined in Eq.~\eqref{eq:reward}, $\gamma \in [0,1]$ is the discount factor and $c_{t}$ is the congestion flag. The congestion flag helps in setting the target model to a lower value for decisions involving congestion, making the model learn better. The model only starts learning once the length of the replay buffer is more than a predefined batch size, denoted as $bSize$ by minimizing the loss function $L(\theta_t)$ using stochastic gradient descent given by:
 \begin{equation}
     L(\theta_t) = \mathbb{E}[(y_{t} - Q_t(s_t, a_{f_t}|\theta_t))^2].
     \label{eq:loss}
 \end{equation}
 %The probability of sampling each transition with priority $P_{f_t}$ is given by:
% \begin{equation}
%     P(t) = \frac{p_{f_t}^\alpha}{\sum_{j=1}^{|B|} p_j^\alpha}
% \end{equation}
 We note that using the target model ($Q^{'}$) leads to more stability in the learning process and helps the DRL framework to learn more effectively. The weights of the target model ($\theta^{'}$) are replaced by that of the main model ($\theta$) after every training episode, e.g., after every $100$ traffic flows.
 
\begin{algorithm}[t]
   \caption{Deep Q-Learning for Adaptive Traffic Routing}
   \label{alg:dqn}
   \begin{algorithmic}[1]
    \Require Network topology 
    \State Initialize $\theta$ and $\theta^{'}$ using Glorot (Xavier) initialization.
    \State $B \leftarrow \{\}$; $count \leftarrow 0$
    \While {\texttt{True}}
        \State Receive $f_t$ = (\texttt{src}, \texttt{dest}, $T^\text{req}$)
        \State Calculate the path $a_{f_t}$ using Eq.~\eqref{eq:cal_policy}.
        \State Enforce $a_{f_t}$ to network and calculate $R_{t}$ using Eq.~\eqref{eq:reward}
        \State Store 
        $\{s_{t}$,$a_{f_t}$,$f_{t}$,$R_{t}$,$s_{t+1}\}$ with priority $p_{f_t}$ into $B$
        \If {$length(B) > bSize$}
        \State Sample a batch from priority experience replay $B$
        \State Calculate priority $p_{f_t}$ using Eq.~\eqref{eq:priority}
        \State Calculate the $q$-value of the target model (Eq.~\eqref{eq:target})   
        \State Update weights of the main model ($\theta$) using Eq.~\eqref{eq:loss}
        \EndIf
    \State $count \leftarrow count + 1$
    \If {$count \mod 100 = 0$}
    \State $\theta^{'} \leftarrow \theta$
    \EndIf
      \EndWhile
   \end{algorithmic}
\end{algorithm}

\section{Experiments}
\label{sec:experiments}

\subsection{Experimental Setting}

We implemented the DRL framework using the OpenAI Gym framework~\cite{brockman2016openai} with  Tensorflow~\cite{abadi2016tensorflow} and StellarGraph~\cite{csiro-data61-stellargraph} libraries. The network topology was modeled using the NetworkX library~\cite{hagberg2008exploring}. Apart from the neural network architecture presented in the previous section, additional hyper-parameters used in the training process are presented in Table~\ref{tab:hyperparameters}. These hyper-parameters were obtained after the initial experiments for tuning and optimization.

\begin{table}[t]
\centering
\small
 \caption{Hyper-parameters}
 \label{tab:hyperparameters}
  \begin{tabular}{ll}
   \toprule
   Parameter & Value \\
   \midrule
    Learning Rate & 0.00025 \\
    Final layer activation function & Softmax \\
    Discount Factor & 0.99 \\
    Weight Initializer & Xavier \\
    Learning Updates per learning session & 10 \\
    Batch size ($bSize$) & 256 \\
    Epsilon Decay & 0.99995 \\
    GCN hidden layers & [128, 128, 128, 64] \\
   \bottomrule
   \end{tabular} 
   \vspace{-2ex}
\end{table}

We evaluated the performance of our framework on two network topologies: a random 9-node topology with $16$ links and NSFNET. NSFNET is a network comprising $14$ nodes, created by the National Science Foundation (NSF) to facilitate academic research collaboration among U.S. universities. For both networks, we connected each node with a single host to generate the traffic. All the links of both networks have a capacity of 100 Mbps. The distance among nodes is randomly generated in the range of $[500,700]$ meters.

We used the end-to-end throughput and end-to-end delay of the traffic flows generated in networks as performance metrics. To evaluate the evolution of these metrics over time, we calculated the metrics for every $100$ traffic flows accommodated and then took the mean for every $10$ values obtained. The delay for each traffic flow is computed as the sum of propagation and transmission delay. For the DRL framework, we additionally measured the evolution of reward in the same manner. We compared our DRL framework implementing DGCNN (denoted as DRL-GCNN) with two baselines.
\begin{itemize}
    \item We used our DRL framework but implemented multi-layer perceptron (MLP) for the neural networks (DRL-MLP).
    \item The OSPF routing algorithm. 
\end{itemize}

We performed various experimental scenarios with different (and dynamic) traffic behaviors to evaluate the performance of the proposed framework and the baselines. We provide a detailed description and analysis of the results below.

\subsection{Analysis of Experimental Results}

\subsubsection{Effectiveness of DGCNN}

\begin{figure}[t]
   \centering
   \subfloat[Random network]{\includegraphics[width=0.24\textwidth]{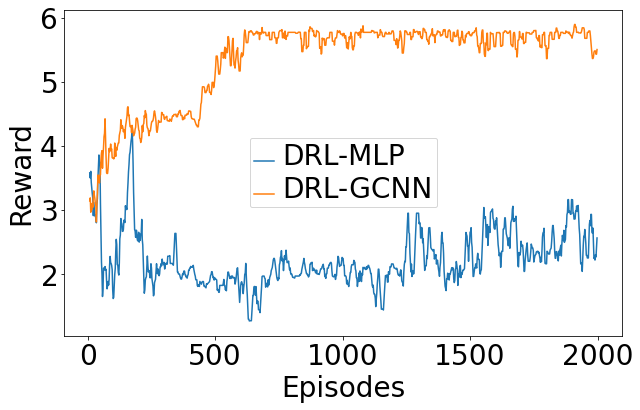}\label{subfig:train_rew_random}}
    %\\
    \subfloat[NSFNET]{\includegraphics[width=0.24\textwidth]{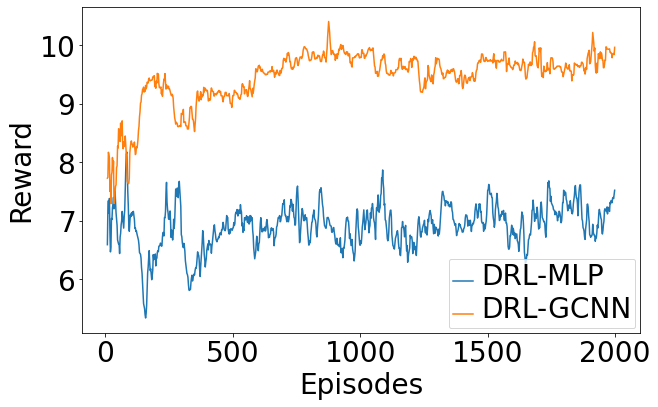}
    \label{subfig:train_rew_nsfnet}}
    \caption{Reward obtained using DGCNN and MLP.}
    \label{fig:train_rew}
    \vspace{-3ex}
\end{figure}

\begin{figure}[t]
   \centering
   \subfloat[Random network]{\includegraphics[width=0.24\textwidth]{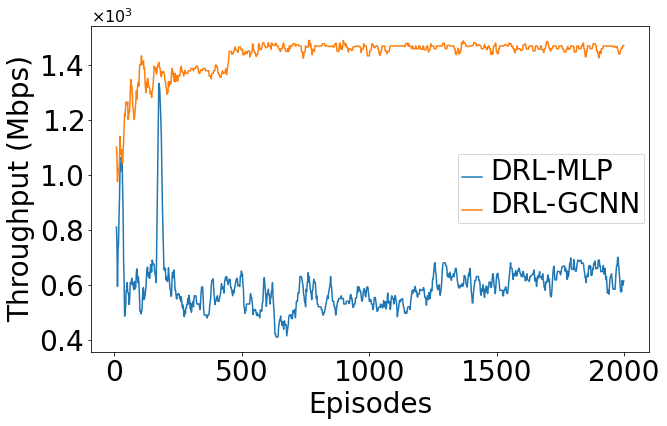}\label{subfig:train_thput_random}}
    %\\
    \subfloat[NSFNET]{\includegraphics[width=0.24\textwidth]{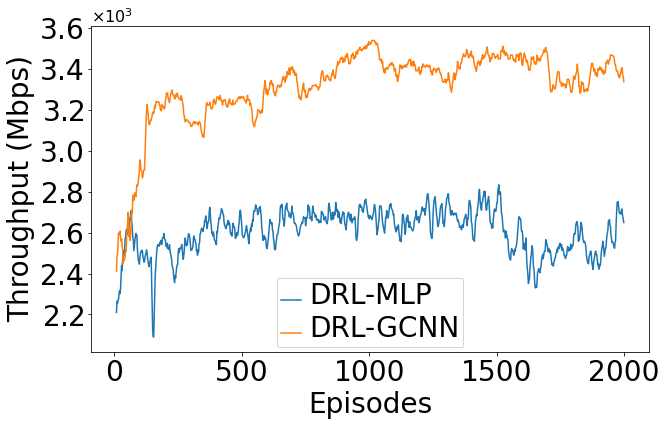}\label{subfig:train_thput_nsfnet}}
    \caption{Throughput obtained using DGCNN and MLP.}
    \label{fig:train_thput}
    \vspace{-2ex}
\end{figure}

\begin{figure}[t]
   \centering
   \subfloat[Random network]{\includegraphics[width=0.24\textwidth]{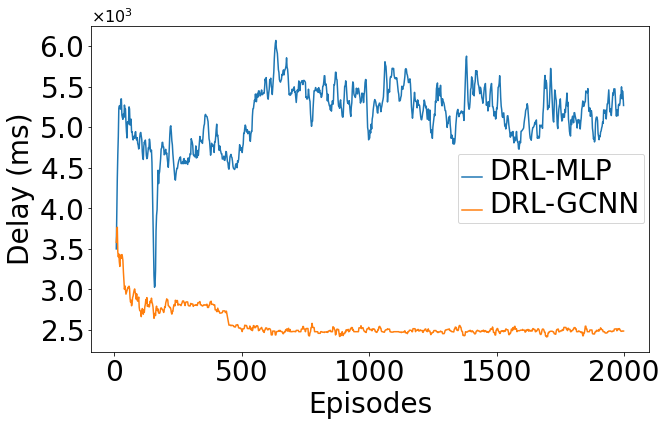}\label{subfig:train_delay_random}}
    %\\
    \subfloat[NSFNET]{\includegraphics[width=0.24\textwidth]{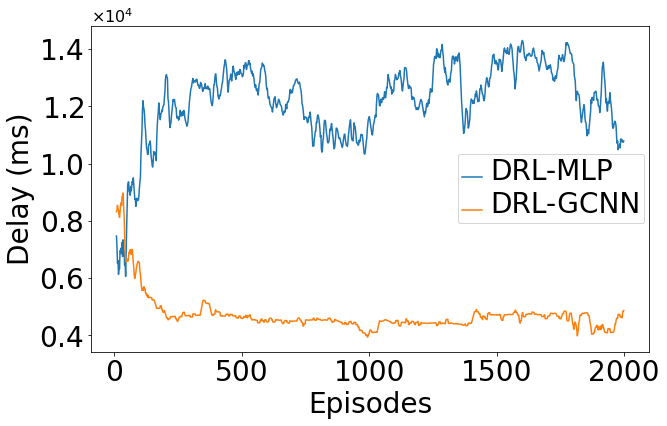}\label{subfig:train_delay_nsfnet}}    
    \caption{Delay incurred using DGCNN and MLP.}
    \label{fig:train_delay}
    \vspace{-2.8ex}
\end{figure}

We demonstrate the effectiveness of DGCNN in learning traffic behavior and making routing decisions over MLP in the DRL framework. For a fair comparison, we just replace DGCNN with MLP in the DRL framework and keep the environment and other parameters the same. For both networks, traffic flows are randomly generated with the requested rate randomly selected in the range of $[40, 60]$ Mbps. Both models learn traffic behavior and make routing decisions for $2000$ episodes, each having $100$ flows. The number of concurrent traffic flows in the network is maintained in the range of $[10,20]$.

In Fig.~\ref{fig:train_rew}, Fig.~\ref{fig:train_thput}, and Fig.~\ref{fig:train_delay}, we present the reward, throughput, and delay on both networks, respectively. It is observed that DRL-GCNN is able to achieve the best possible solution by maximizing the reward, leading to increased network throughput and decreased network delay. After the first $500$ episodes of learning and exploring traffic behavior, DRL-GCNN is able to stabilize its performance. It is also observed that DRL-MLP fails to solve the route optimization problem and remains confined to a local optimal solution with a lower reward even after training for $2000$ episodes. This is due to the lack of topology knowledge as well as large action space. It is worth mentioning that the larger the network, the harder the learning of the DRL framework, thus the more the fluctuation of reward, throughput, and delay. 

\subsubsection{Adaptiveness of DRL-GCNN}

In this experiment, we demonstrate the adaptiveness of DRL-GCNN against traffic dynamics. We run the DRL-GCNN with $4000$ episodes with two traffic variation scenarios:
\begin{itemize}
    \item Decreasing traffic demand: In the first $2000$ episodes, traffic flows are generated with the requested rate randomly selected in the range of $[40,60]$ Mbps and reduced to $[20,40]$ Mbps in the remaining $2000$ episodes.
    \item Increasing traffic congestion: We keep the same requested rate ($[40,60]$ Mbps) but we increase the number of concurrent traffic flows in the second half of the experiment.
\end{itemize}

\begin{figure}[t]
   \centering
   \subfloat[Decreasing Traffic Demand]{\includegraphics[width=0.24\textwidth]{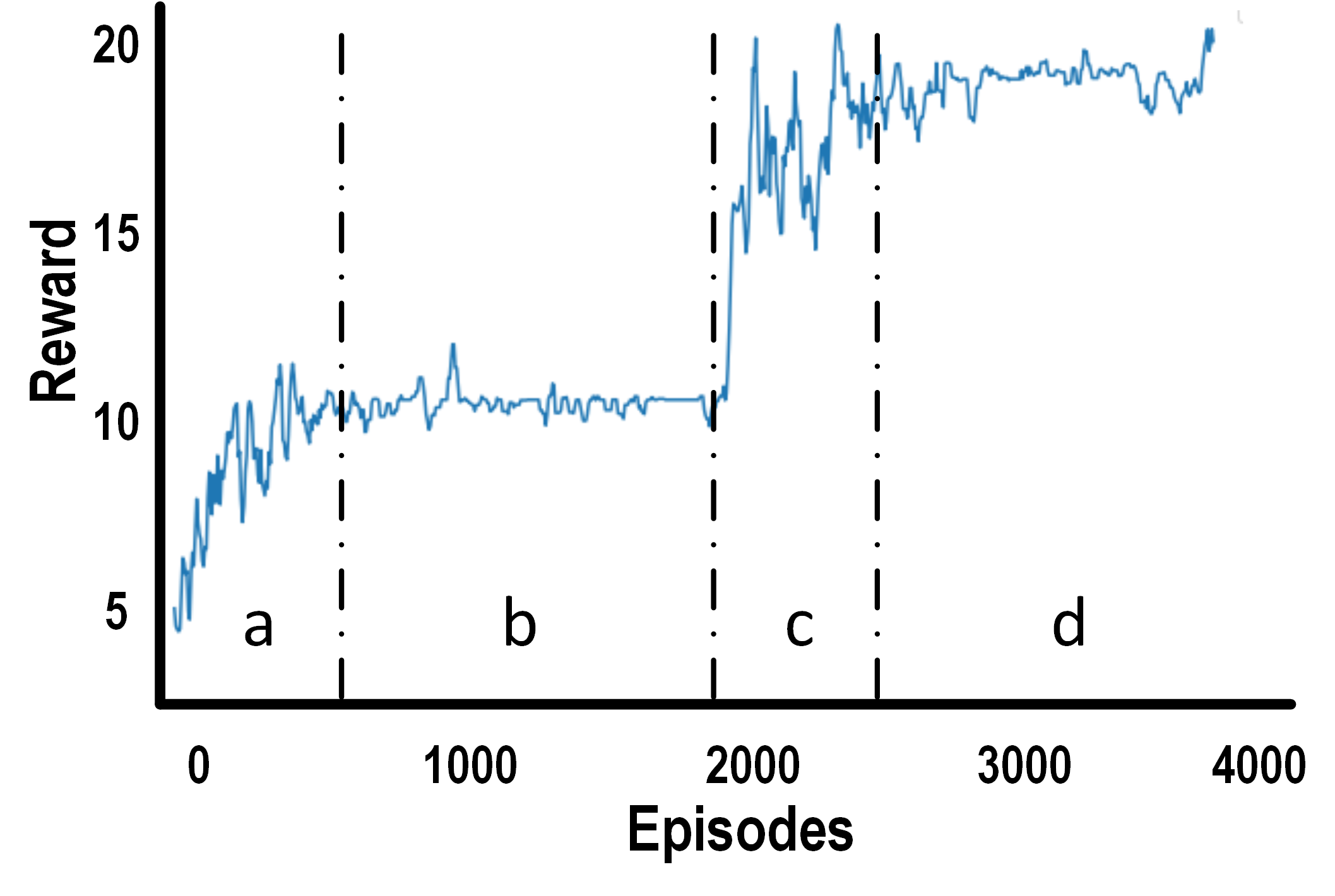}\label{subfig:traffic1_rew}}
    %\\
    \subfloat[Increasing Traffic Congestion]{\includegraphics[width=0.24\textwidth]{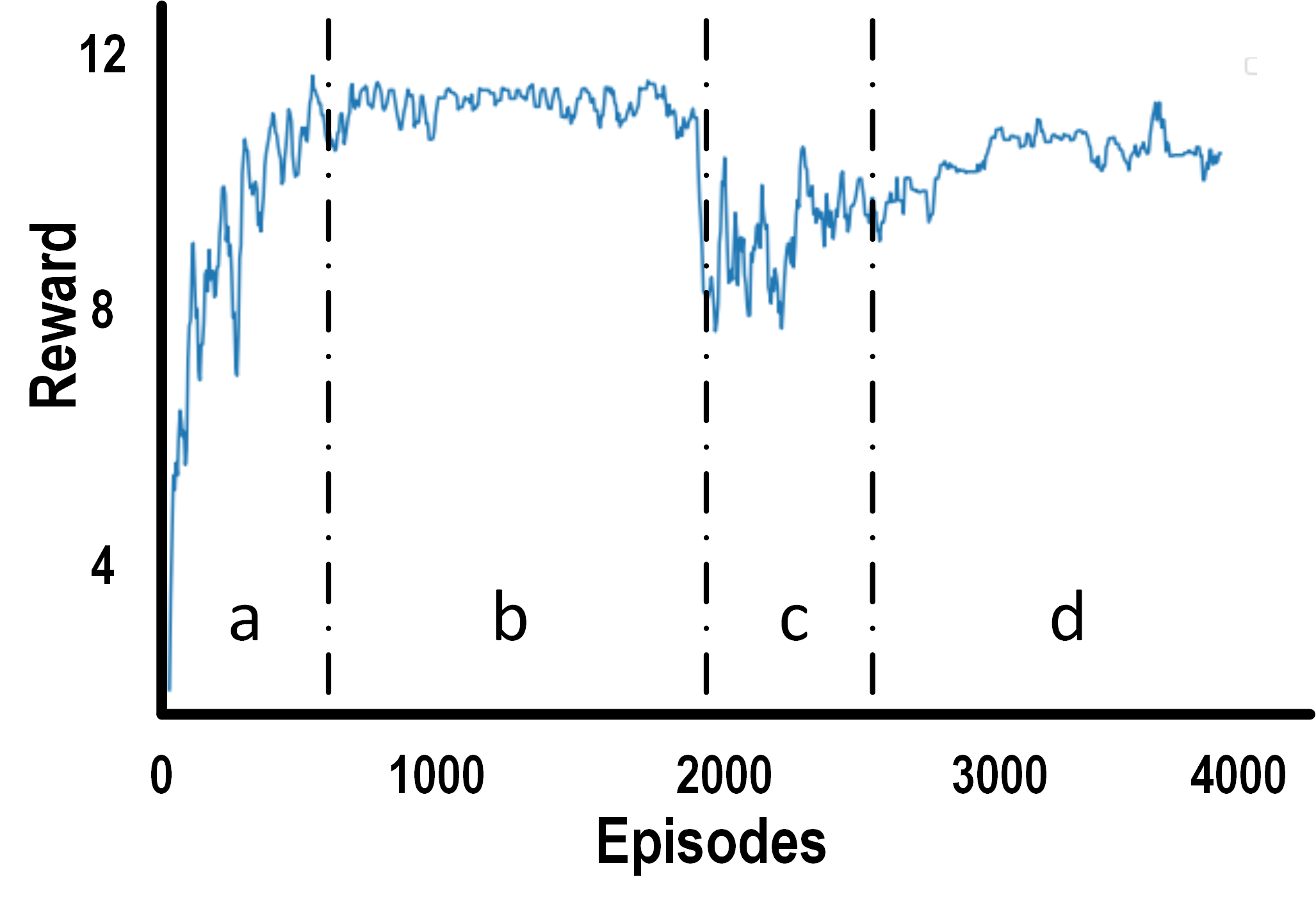}
    \label{subfig:train_rew_nsfnet}}
    \caption{Reward obtained with traffic variations.}
    \label{fig:test_traffic_rew}
    \vspace{-2.8ex}
\end{figure}

\begin{figure}[t]
   \centering
   \subfloat[Decreasing Traffic Demand]{\includegraphics[width=0.24\textwidth]{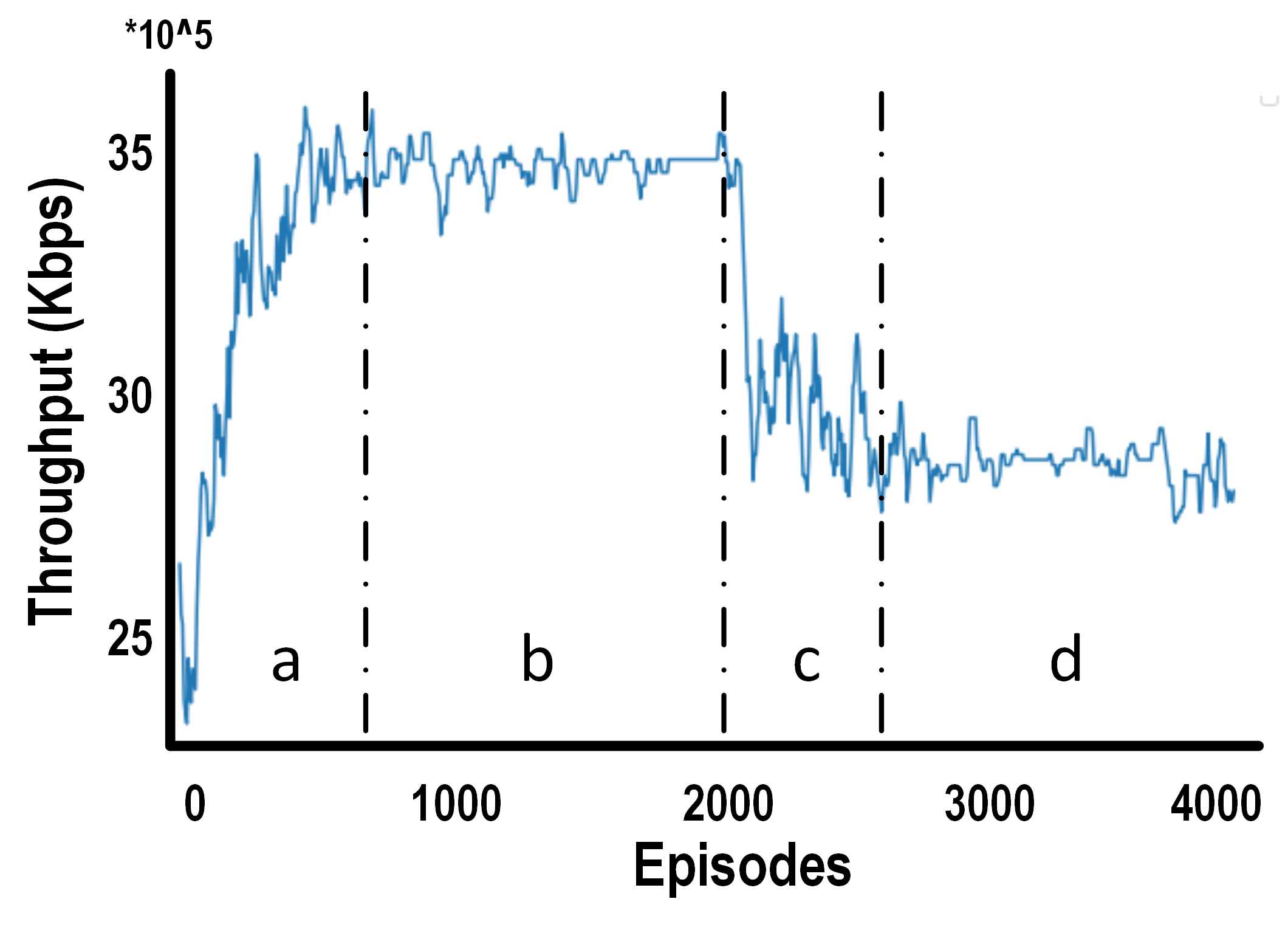}\label{subfig:traffic1_thput}}
    %\hfill
    \subfloat[Increasing Traffic Congestion]{\includegraphics[width=0.24\textwidth]{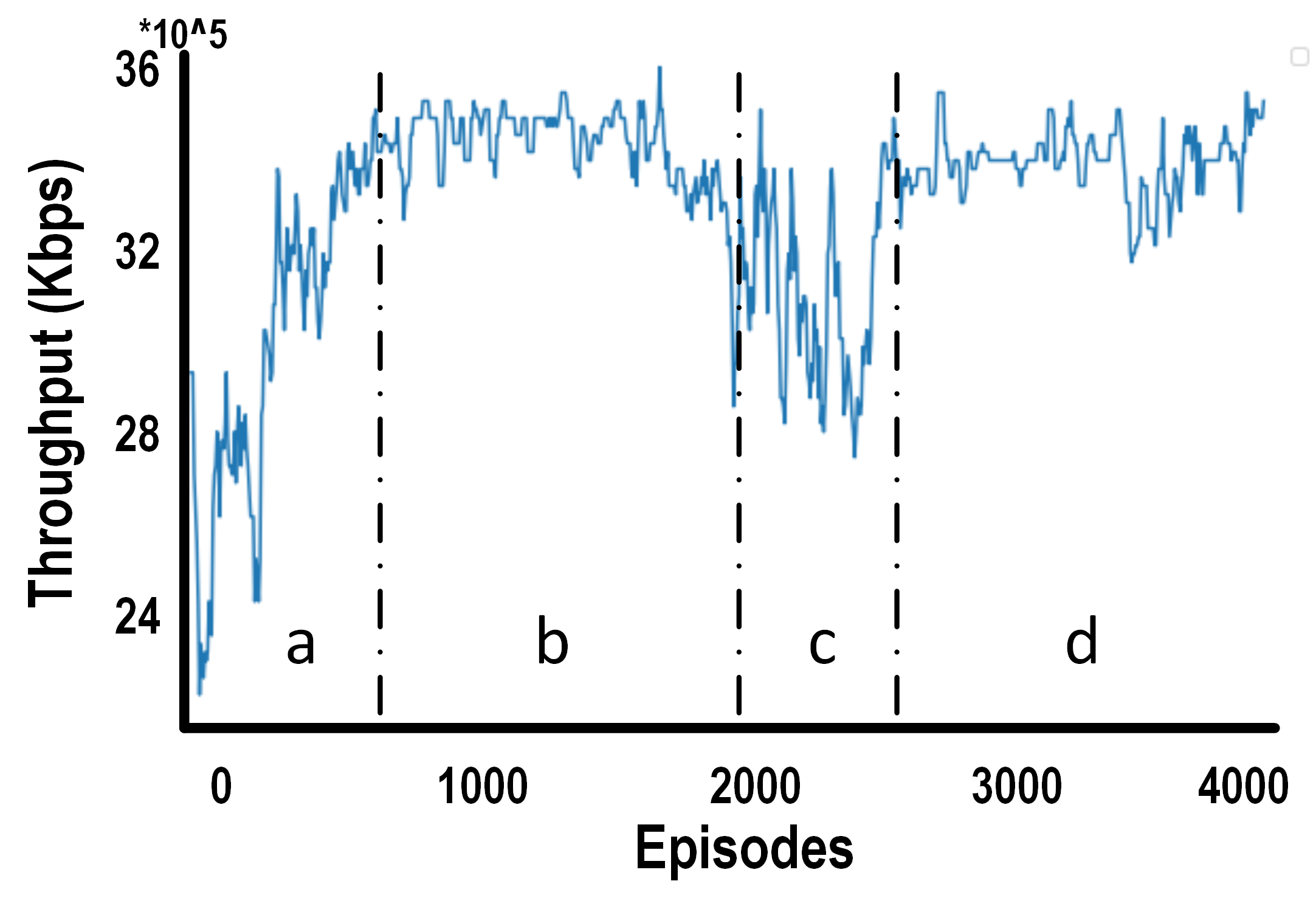}\label{subfig:traffic2_thput}}
    \caption{Throughput obtained with traffic variations.}
    \label{fig:test_traffic_thput}
    \vspace{-2.5ex}
\end{figure}

\begin{figure}[t]
   \centering
   \subfloat[Decreasing Traffic Demand]{\includegraphics[width=0.24\textwidth]{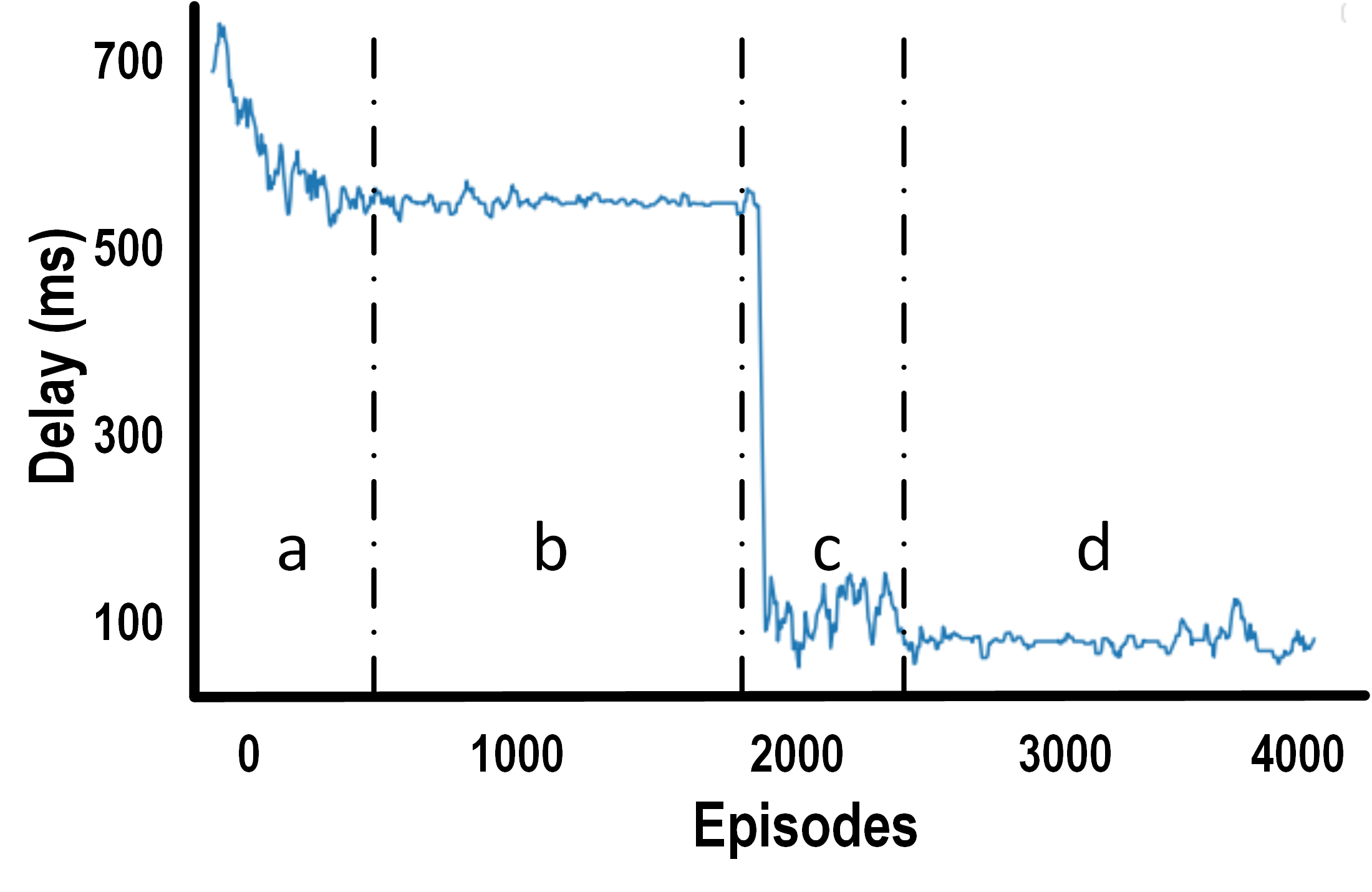}\label{subfig:traffic1_delay}}
    %\hfill
    \subfloat[Increasing Traffic Congestion]{\includegraphics[width=0.24\textwidth]{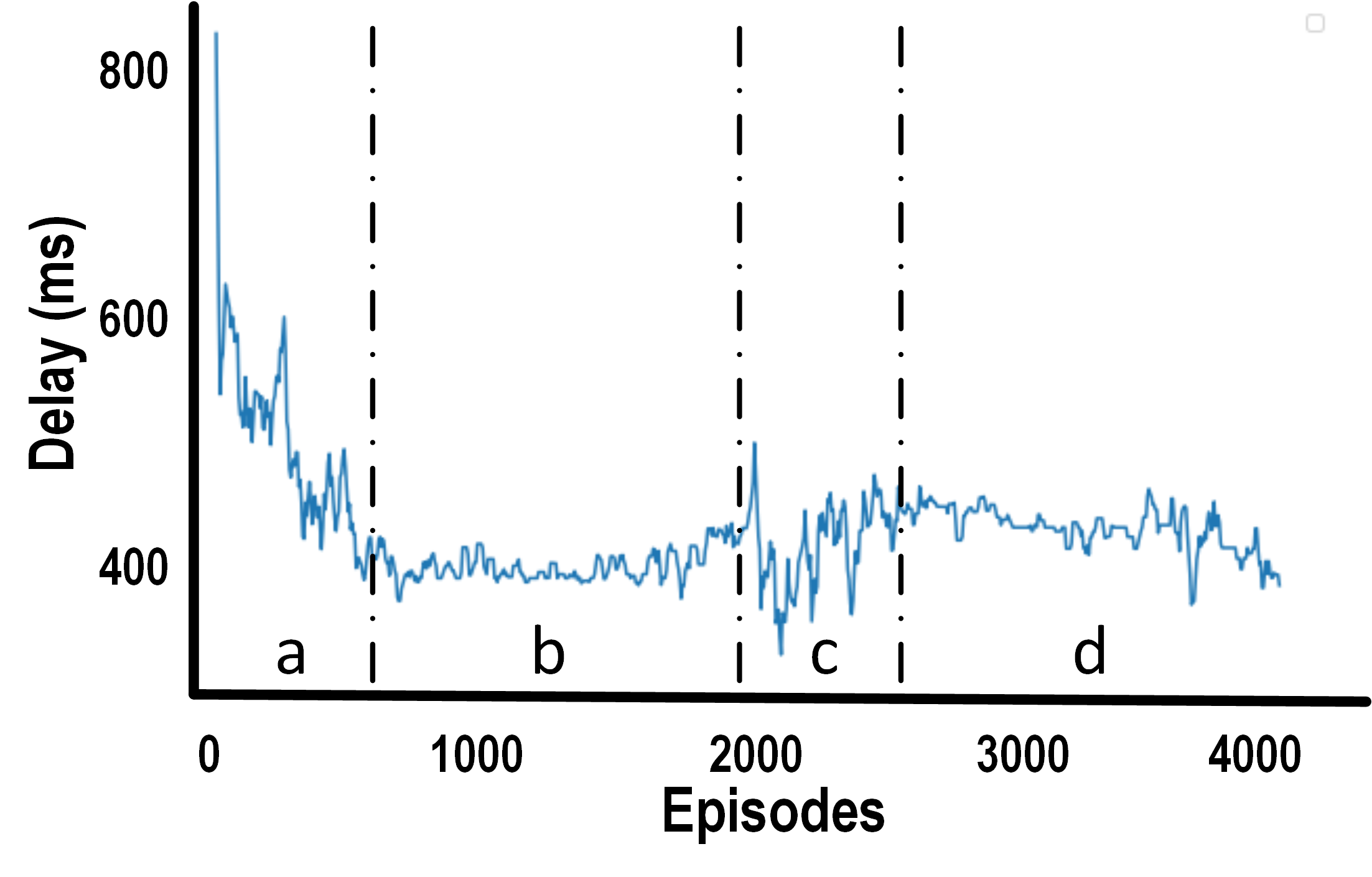}\label{subfig:traffic2_delay}}    
    \caption{Delay incurred with traffic variations.}
    \label{fig:test_traffic_delay}
    \vspace{-2.5ex}
\end{figure}

In Fig.~\ref{fig:test_traffic_rew}, Fig.~\ref{fig:test_traffic_thput}, and Fig.~\ref{fig:test_traffic_delay}, we present the obtained results with additional annotations. It is observed that DRL-GCNN is robust against traffic fluctuations and can quickly adapt to new traffic behavior by updating the DGCNN models used in the framework. The framework starts with an exploration to learn traffic behavior (Region \texttt{a}) and then stabilizes its performance before entering the exploration stage again (Region \texttt{c}) on encountering the changes in traffic behavior. The convergence speed of the framework is dependent upon the frequency of gathering network statistics. Approximately 500 epochs are required for adaptation to the updated traffic pattern. In the scenario of decreasing traffic demand, the corresponding decrease in throughput and delay are observed due to the decrease in traffic load (Fig.~\ref{subfig:traffic1_thput} and Fig.~\ref{subfig:traffic1_delay}). As expected, in the scenario of increasing traffic congestion, there is a slight decrease in throughput (Fig.~\ref{subfig:traffic2_thput}) and an increase in delay (Fig.~\ref{subfig:traffic2_delay}), leading to decreased reward (Fig.~\ref{subfig:train_rew_nsfnet}).

\subsubsection{Comparison with Existing Works}

\begin{figure}[t]
    \centering
    \subfloat[Throughput]{\includegraphics[width=0.24\textwidth]{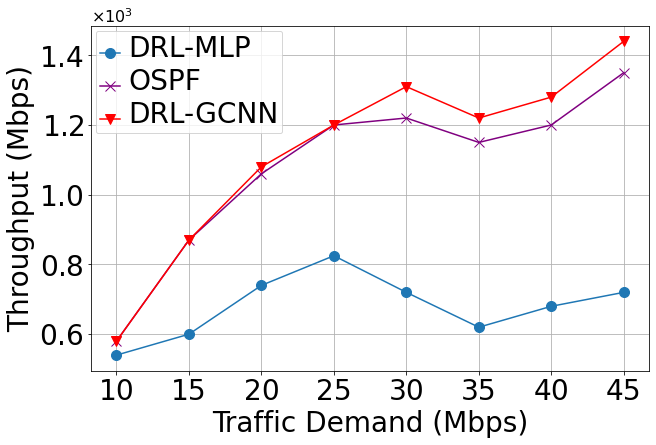}\label{subfig:thput_test_random}}
    \subfloat[Delay]{\includegraphics[width=0.24\textwidth]{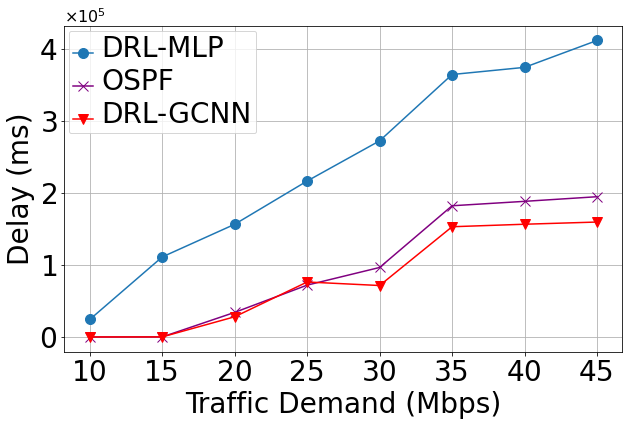}\label{subfig:delay_test_random}}
  \caption{Performance on random network with increasing traffic.}
   \label{fig:random_test}
   \vspace{-3ex}
\end{figure}
\begin{figure}[t]
    \centering
    \subfloat[Throughput]{\includegraphics[width=0.24\textwidth]{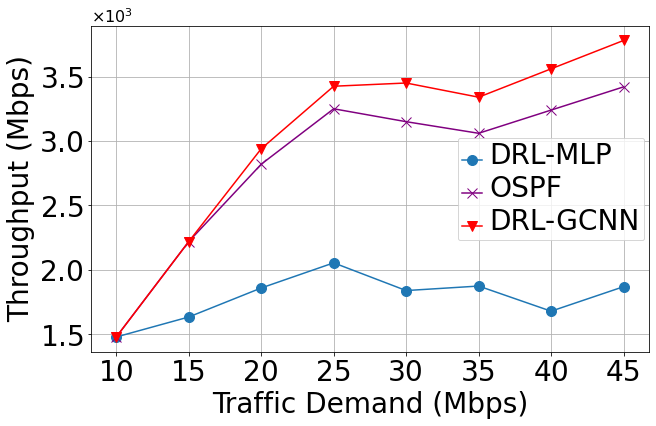}\label{subfig:thput_test_nsfnet}}
    \subfloat[Delay]{\includegraphics[width=0.24\textwidth]{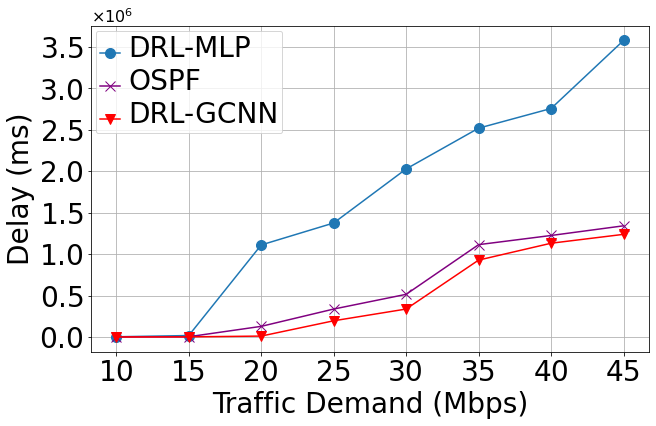}\label{subfig:delay_test_nsfnet}} 
   \caption{Performance on NSFNET with increasing traffic.}
   \label{fig:nsfnet_test}
    \vspace{-3ex}
\end{figure}
In this experiment, we freeze the DGCNN model after $2000$ training episodes with traffic flows generated with the rate in the range of $[40,60]$ Mbps and the number of concurrent flows in the range of $[10,20]$. The obtained DGCNN model is then used to test against traffic variations without being able to update the model along with the changes in traffic behavior. The rate of traffic flows is gradually increased after every $10$ episodes. From Fig.~\ref{subfig:thput_test_random} and Fig.~\ref{subfig:thput_test_nsfnet}, we observe that DRL-GCNN increases the network throughput by $5.1\%$ and $7.8\%$ compared to OSPF and by $39.4\%$ and $69.2\%$ compared to DRL-MLP, for both topologies, respectively. On the delay,  Fig.~\ref{subfig:delay_test_random} and Fig.~\ref{subfig:delay_test_nsfnet} show a significant reduction in the network delay by $16.1\%$ and $17.5\%$ compared to OSPF and by $66.5\%$ and $71.2\%$ compared to DRL-MLP on both topologies. Note that the x-axis represents the central values of the corresponding traffic rate of all flows in the examined episodes. Initially, with low traffic demand, OSPF and DRL-GCNN have similar performance. This is due to the fact that with no congestion in the network topology, the shortest path routing is the best solution. However, as the traffic increases, DRL-GCNN performs better in terms of both network throughput and delay in both topologies by avoiding the routes with congestion.

It is to be noted that while the delay increases along with the increase in traffic demand due to heavier traffic congestion, the throughput also increases until attaining the maximum accommodation capacity of the network. Nevertheless, this experiment demonstrates the superiority of the proposed DRL framework over the existing routing methods. 
%It can also be observed that no matter what method is used and no matter which network topology is chosen, both the network throughput and delay go up with the traffic demand. This is because the higher the traffic load, the higher is the throughput, at the same time higher is the delay due to longer waiting time or congestion. Moreover, it is to be noted that the throughput does not increase monotonically because higher traffic demands may even lead to poorer throughput due to congestion and packet losses when the network becomes saturated. It is to be noted that the DRL-GCNN is also robust against changes in traffic load and network topology since it performs consistently better than both the methods across different traffic demand settings of both topologies. 

\section{Conclusion}
\label{sec:conclusion}

In this paper, we presented a deep reinforcement learning framework using deep graph convolutional neural networks (DRL-GCNN) for adaptive traffic routing. We proposed to capture traffic features at both link and node levels and combine them with network topology to form a graph-based data structure to train the deep graph convolutional neural network, integrated in the DRL framework for path decision. We developed a training algorithm that allows the DRL framework to switch between the exploration phase and the exploitation phase, thus achieving a fast model updating and quickly adapting to changes in traffic behavior. We performed extensive experiments with various scenarios and compared the performance of the proposed framework with existing solutions. The experimental results confirm the effectiveness and adaptiveness of the proposed framework by increasing the network throughput by up to $7.8\%$ and reducing the traffic delay by up to $16.1\%$ compared to OSPF. For future work, we will explore the integration of an SDN controller into a real experimental testbed to capture real-time network statistics for training our DRL framework and evaluate its performance. We will also investigate the impact of other DRL techniques on the performance of the proposed framework.

\section*{Acknowledgment}
This research is supported by the Ministry of Education, Singapore, under its Academic Research Tier 1 (Grant WBS number: R-R12-A405-0003).

\section*{Disclaimer}
Any opinions, findings, conclusions, or recommendations expressed in this material are those of the author(s) and do not reflect the views of the Ministry of Education, Singapore.

\bibliographystyle{IEEEtran}
\bibliography{icc2024.bib}

\begin{thebibliography}{10}
\providecommand{\url}[1]{#1}
\csname url@samestyle\endcsname
\providecommand{\newblock}{\relax}
\providecommand{\bibinfo}[2]{#2}
\providecommand{\BIBentrySTDinterwordspacing}{\spaceskip=0pt\relax}
\providecommand{\BIBentryALTinterwordstretchfactor}{4}
\providecommand{\BIBentryALTinterwordspacing}{\spaceskip=\fontdimen2\font plus
\BIBentryALTinterwordstretchfactor\fontdimen3\font minus
  \fontdimen4\font\relax}
\providecommand{\BIBforeignlanguage}[2]{{%
\expandafter\ifx\csname l@#1\endcsname\relax
\typeout{** WARNING: IEEEtran.bst: No hyphenation pattern has been}%
\typeout{** loaded for the language `#1'. Using the pattern for}%
\typeout{** the default language instead.}%
\else
\language=\csname l@#1\endcsname
\fi
#2}}
\providecommand{\BIBdecl}{\relax}
\BIBdecl

\bibitem{intel}
\BIBentryALTinterwordspacing
Intel tofino. [Online]. Available:
  \url{www.intel.sg/content/www/xa/en/products/network-io/programmable-ethernet-switch/tofino-2-series.html}
\BIBentrySTDinterwordspacing

\bibitem{nvidia}
\BIBentryALTinterwordspacing
Nvidia. [Online]. Available:
  \url{https://www.nvidia.com/en-sg/networking/ethernet-switching/cumulus-vx/}
\BIBentrySTDinterwordspacing

\bibitem{jiang2021machine}
H.~Jiang, Q.~Li, Y.~Jiang, G.~Shen, R.~Sinnott, C.~Tian, and M.~Xu, ``When
  machine learning meets congestion control: A survey and comparison,''
  \emph{Computer Networks}, vol. 192, p. 108033, 2021.

\bibitem{tram2019ml}
T.~Truong-Huu, P.~Prathap, P.~M. Mohan, and M.~Gurusamy, ``{Fast and Adaptive
  Failure Recovery using Machine Learning in Software Defined Networks},'' in
  \emph{2019 IEEE ICC Workshops}, 2019.

\bibitem{sutton2018reinforcement}
R.~S. Sutton and A.~G. Barto, \emph{{Reinforcement Learning: An
  Introduction}}.\hskip 1em plus 0.5em minus 0.4em\relax MIT Press, 2018.

\bibitem{valadarsky2017learning}
A.~Valadarsky, M.~Schapira, D.~Shahaf, and A.~Tamar, ``{Learning to Route},''
  in \emph{ACM HotNets 2017}, Nov. 2017.

\bibitem{wang2017machine}
M.~Wang, Y.~Cui, X.~Wang, S.~Xiao, and J.~Jiang, ``Machine learning for
  networking: Workflow, advances, and opportunities,'' \emph{IEEE Network},
  vol.~32, no.~2, pp. 92--99, 2017.

\bibitem{zhou2018graph}
J.~Zhou, G.~Cui, Z.~Zhang, C.~Yang, Z.~Liu, and M.~Sun, ``{Graph Neural
  Networks: A Review of Methods and Applications},'' \emph{CoRR}, 2018.

\bibitem{xu2018experience}
Z.~Xu, J.~Tang, J.~Meng, W.~Zhang, Y.~Wang, C.~H. Liu, and D.~Yang,
  ``{Experience-Driven Networking: A Deep Reinforcement Learning-Based
  Approach},'' in \emph{IEEE INFOCOM 2018}, Apr. 2018.

\bibitem{abbasloo2020classic}
S.~Abbasloo, C.-Y. Yen, and H.~J. Chao, ``{Classic Meets Modern: A Pragmatic
  Learning-Based Congestion Control for the Internet},'' in \emph{ACM SIGCOMM
  '20}, 2020.

\bibitem{lan2019deep}
D.~Lan \emph{et~al.}, ``{A Deep Reinforcement Learning-Based Congestion Control
  Mechanism for NDN},'' in \emph{IEEE ICC 2019}, 2019.

\bibitem{emara2020eagle}
S.~Emara, B.~Li, and Y.~Chen, ``{Eagle: Refining Congestion Control by Learning
  from the Experts},'' in \emph{IEEE INFOCOM 2020}, 2020.

\bibitem{roshdi2021deep}
M.~Roshdi, S.~Bhadauria, K.~Hassan, and G.~Fischer, ``{Deep Reinforcement
  Learning based Congestion Control for V2X Communication},'' in \emph{IEEE
  PIMRC 2021}, 2021.

\bibitem{chen2021ran}
M.~Chen, R.~Li, Z.~Zhao, and H.~Zhang, ``{RAN Information-assisted TCP
  Congestion Control via DRL with Reward Redistribution},'' in \emph{IEEE ICC
  2021 Workshops}, 2021.

\bibitem{emara2022pareto}
S.~Emara, F.~Wang, B.~Li, and T.~Zeyl, ``Pareto: Fair congestion control with
  online reinforcement learning,'' \emph{IEEE Trans. Netw. Sci. Eng.}, vol.~9,
  no.~5, pp. 3731--3748, Sept.-Oct. 2022.

\bibitem{almasan2022deep}
P.~Almasan \emph{et~al.}, ``{Deep reinforcement learning meets graph neural
  networks: Exploring a routing optimization use case},'' \emph{Computer
  Communications}, vol. 196, pp. 184--194, 2022.

\bibitem{bhavanasi2022routing}
S.~S. Bhavanasi, L.~Pappone, and F.~Esposito, ``Routing with graph
  convolutional networks and multi-agent deep reinforcement learning,'' in
  \emph{2022 IEEE Conference on Network Function Virtualization and Software
  Defined Networks (NFV-SDN)}.\hskip 1em plus 0.5em minus 0.4em\relax IEEE,
  2022, pp. 72--77.

\bibitem{yang2022joint}
L.~Yang, Y.~Wei, F.~R. Yu, and Z.~Han, ``Joint routing and scheduling
  optimization in time-sensitive networks using
  graph-convolutional-network-based deep reinforcement learning,'' \emph{IEEE
  Internet of Things Journal}, vol.~9, no.~23, pp. 23\,981--23\,994, 2022.

\bibitem{huang2022deep}
R.~Huang, W.~Guan, G.~Zhai, J.~He, and X.~Chu, ``Deep graph reinforcement
  learning based intelligent traffic routing control for software-defined
  wireless sensor networks,'' \emph{Applied Sciences}, vol.~12, no.~4, p. 1951,
  2022.

\bibitem{zhang2018endtoend}
M.~Zhang \emph{et~al.}, ``{An End-to-End Deep Learning Architecture for Graph
  Classification},'' in \emph{AAAI-18}, New Orleans, USA, 2018.

\bibitem{mnih2013playing}
V.~Mnih, K.~Kavukcuoglu, D.~Silver, A.~Graves, I.~Antonoglou, D.~Wierstra, and
  M.~Riedmiller, ``{Playing Atari with Deep Reinforcement Learning},'' in
  \emph{NIPS Deep Learning Workshop 2013}, 2013.

\bibitem{brockman2016openai}
G.~Brockman \emph{et~al.}, ``{OpenAI Gym},'' \emph{arXiv}, 2016.

\bibitem{abadi2016tensorflow}
M.~Abadi \emph{et~al.}, ``{TensorFlow: A System for Large-Scale Machine
  Learning},'' in \emph{USENIX OSDI 2016}, 2016.

\bibitem{csiro-data61-stellargraph}
\BIBentryALTinterwordspacing
{CSIRO's Data61}, ``Stellargraph machine learning library,'' 2018. [Online].
  Available: \url{https://github.com/stellargraph/stellargraph}
\BIBentrySTDinterwordspacing

\bibitem{hagberg2008exploring}
A.~A. Hagberg, D.~A. Schult, and P.~J. Swart, ``Exploring network structure,
  dynamics, and function using networkx,'' in \emph{Proc. 7th Python in Science
  Conf.}, G.~Varoquaux, T.~Vaught, and K.~J. Millman, Eds., 2008.

\end{thebibliography}

\end{document}